\documentclass[showpacs,showkeys,amsmath,amssymb,preprint,nofootinbib]{revtex4}
\usepackage{graphicx}
\usepackage{epsfig}
\usepackage{bm}
\usepackage{float}
\usepackage[T1]{fontenc}
\usepackage[utf8]{inputenc}
\usepackage{graphicx}
\usepackage{bm}
\usepackage{amsmath}
\usepackage{xcolor}
\DeclareUnicodeCharacter{2212}{\textendash}
\def\beq{\begin{eqnarray}}

\def\eeq{\end{eqnarray}}

\begin{document}
\title[HDET]{Holographic dark energy in curved spacetime for interacting fluids}

\author{Azucena Bola\~nos$^{a,b}$}
\email{azucena.bolanos@iberopuebla.mx}

\author{Miguel Cruz$^{c}$}
\email{miguelcruz02@uv.mx}

\author{Samuel Lepe$^{d}$}
\email{samuel.lepe@pucv.cl}

\author{David Rogelio M\'arquez-Castillo$^{c,e}$}
\email{david.marquez@cinvestav.mx}

\affiliation{$^a$Tecnol\'ogico de Monterrey, Department of Science Campus Puebla, Av. Atlixc\'ayotl 2301, Puebla, M\'exico\\
$^b$Departamento de Ciencias e Ingenier\'\i as, Universidad Iberoamericana, Boulevard del Ni\~no Poblano 2901, Reserva Territorial Atlixc\'ayotl, San Andr\'es Cholula, Puebla, M\'exico\\
$^c$Facultad de F\'\i sica, Universidad Veracruzana 91097, Xalapa, Veracruz, M\'exico\\
$^d$Instituto de F\'\i sica, Facultad de Ciencias, Pontificia Universidad Cat\'olica de Valpara\'\i so, Av. Brasil 2950, Valpara\'\i so, Chile\\
$^e$Departamento de F\'\i sica, CINVESTAV-IPN, Av. Instituto Polit\'ecnico Nacional 2508, Col. San Pedro Zacatenco, 07360, Gustavo A. Madero, CDMX, M\'exico}

\date{\today}

\begin{abstract}
In this work we explore the scope of two holographic dark energy models within the interacting scenario for the dark sector and spatial curvature. For one holographic model we consider the usual formula for the dark energy density with the Hubble scale and the second model is given in terms of a function instead of a constant parameter as in the usual holographic formula. In this description both holographic models admit a future singularity. The use of recent cosmological data is considered only to support the physical results of our proposal. The interaction term for each holographic model, $Q$, keeps positive along the cosmic evolution and are not given a priori, they are reconstructed from the dynamics of the model. The temperatures for the components of the dark sector are computed and exhibit a growing behavior in both scenarios. The cosmic evolution in this context it is not adiabatic and the second law it is fulfilled only under certain well-established conditions for the temperatures of the cosmic components and positive $Q$-terms.
\end{abstract}


\maketitle
\section{Introduction}
\label{sec:intro}

In contemporary cosmology dark matter and dark energy are two unclear concepts from which we depend to construct the current picture of the observable Universe. This has been a source of motivation to explore some extensions of General Relativity; the cosmic effects of dark energy can be obtained in a minimal modification of Einstein's theory which is usually known as, $\Lambda$CDM model, and consists in the inclusion of a constant term at the gravitational action. However, some issues have been unveiled on the grounds of this theoretical framework suggesting that a dynamical behavior for dark energy seems to be more appropriate to describe the late times behavior of the Universe, see for instance Ref. \cite{jje}. Besides, with the advent of more precise cosmological observations some discrepancies appear with different data sets in the fundamental constant that characterizes the expansion rate of the Universe when the $\Lambda$CDM model is considered to describe the cosmic evolution, this is known as the $H_{0}$ tension, see Ref. \cite{tension} for a general perspective on the different values of the expansion rate coming from different data sets. In this sense, the necessity of cosmological models beyond the standard one is unavoidable. An example of a modified gravity model considered to solve the aforementioned tension can be found in \cite{viscodint}. Other proposal can be found in Ref. \cite{spin}, a non vanishing spin tensor for dark matter seems to provide an alleviation for the aforementioned $H_{0}$ tension. However, a complete review of viable cosmological candidates that could solve the $H_{0}$ tension is given in \cite{catalogue}. On the other hand, since the nature of the dark matter and dark energy is still unsolved, it is possible to consider that both components are allowed to interact to each other, or in other words, an interchange of energy could exist between such components. In general, this interchange of energy is mediated by the existence of a $Q$ interaction term that is introduced in the continuity equations of the energy densities of both cosmic components. It is worthy to mention that the interaction scheme has been studied widely in the literature and it is not ruled out by the recent cosmological observations, an interesting review on the topic can be found in Ref. \cite{interact}. In fact, a reduction of the $H_{0}$ tension to 2.5$\sigma$ was reported recently within the interacting scenario \cite{tensioninteract}, this is an improvement on the tensions reported in \cite{tension}, which variate between 3.1 and 5.8$\sigma$. See also Ref. \cite{wimp}, where the interaction of WIMP dark matter with dark energy is considered.\\

From the consideration of some heuristic arguments used in particle physics emerges an interesting proposal to deal with the dark energy problem known as holographic principle \cite{principle}, the description of a physical system within a volume can be performed using only the degrees of freedom residing on the boundary. Using this scenario, it was possible to establish an upper bound for the entropy, $S$, in a region of size $L$, $S \leq M^{2}_{P}L^{2}$, where $M_{P}$ is the Planck mass, as can be seen at large scales such quantum field theory breaks down. Therefore the energy contained in a region of volume $L^{3}$ should not exceed the energy of a black hole of the same size, i.e, $L^{3}\lambda^{3}_{UV} \leq S_{BH} = M^{2}_{P}L^{2}$, where $\lambda_{UV}$ is identified as the ultraviolet cut-off (vacuum energy) and $L$ acts as the infrared cut-off (scale of the Universe or characteristic length) which scales as $\lambda^{-3}_{UV}$, both quantities are related. In order to avoid the saturation of the aforementioned expression for the energy we can consider that $L$ scales as $\lambda^{-2}_{UV}$, leading to $L^{3}\lambda^{4}_{UV} \leq M^{2}_{P}L$. At cosmological level, the idea behind this approach is to have a theoretical model that reproduces the amount of dark energy in the Universe, as obtained from cosmological observations. This was developed by Li in Ref. \cite{li} by identifying $\lambda^{4}_{UV}$ as the holographic dark energy and $L$ as the Hubble scale $H^{-1}$, the Li model was tested with observations in Ref. \cite{plhde}. Of course, the election of the characteristic length is not unique, see for instance Refs. \cite{holo1, holo18, holo2, holo3, holo4, npb1, npb2, odintsov, odintsovde, holo5, holobs}, where some different forms of $L$ and/or generalizations of the holographic principle can be found. In Ref. \cite{holo6} the holographic dark energy was studied in the context of the interacting dark sector. An interesting scenario for holographic dark energy is explored in Ref. \cite{holo7}, where the matter production effects are taken into account, such effects contribute on the cosmic expansion since are characterized by a negative effective pressure. The holographic principle also supports the physics of the early Universe, in Ref. \cite{Nojiri:2019kkp} an inflationary model based on the holographic approach is studied and in Ref. \cite{victorhp} the inflation process can be understood naturally by considering some holographic arguments.\\ 

Without question we live in an Universe for which a plethora of cosmological models could be adapted. However, we must pay attention to the hints revealed by the astrophysical observations, some of the tensions found in modern cosmology allow small deviations from a flat configuration \cite{snowmass}, then the understanding of the effects of curvature parameter on the cosmic evolution is crucial. See for instance Ref. \cite{pavonk}, where it was found that the inclusion of spatial curvature leads to consistent results at thermodynamics level, i.e., its consideration does not alter the second law.\\ 

This work is organized as follows: in Section \ref{sec:background} is provided a brief description of the background dynamics for the interacting scheme plus spatial curvature. The basic dynamical equations are established and a general expression for the dark energy parameter state is given in terms of some relevant cosmological quantities. Section \ref{sec:cuts} is devoted to construct the interacting scenario when a holographic approach is considered for the dark energy component. We consider two different holographic models and we label them as Model I and Model II, being the latter a generalization of the usual holographic formula, we replace the holographic constant parameter $c$ by a function $c(z)$, which origin is inspired in the apparent horizon, as we will see below, both holographic models exhibit a future singularity. The general behavior of the interaction term in both models is also shown using the best fit values obtained from the statistical analysis discussed in Appendix \ref{sec:stat}. Section \ref{sec:thermo} is oriented to the thermodynamics discussion of these holographic scenarios. By means of the effective temperature method we compute the temperatures of the cosmic components, using the constrained values for the cosmological parameters we show that such temperatures are well defined in both cases studied and the second law will be fulfilled in the future evolution of the cosmic expansion. In Section \ref{sec:final} we give the final comments of our work. In this work we will consider $8\pi G=c=k_{B}=1$ units.   

\section{Interacting dark sector plus curvature}
\label{sec:background}
In this section we will find a brief discussion of some results explored previously by part of the authors in Ref. \cite{cruz18} for the interacting dark sector with spatial curvature contribution. If we consider a curved Universe filled with dark matter $(\mathrm{m})$ and dark energy $(\mathrm{de})$, the Friedmann constraint can be written in terms of the cosmological redshift, $z$, as follows
\begin{equation}
    E^{2}(z) = \frac{1}{3H^{2}_{0}}(\rho_{\mathrm{m}}(z)+\rho_{\mathrm{de}}(z)) + \Omega_{\mathrm{k}}(z),
    \label{eq:fried}
\end{equation}
where we have considered the normalized Hubble parameter, $E(z) := H(z)/H_{0}$, being $H_{0}$ the value of the Hubble parameter at present time, i.e., $H(z=0)=H_{0}$. On the other hand, we have defined the fractional energy density for curvature as $\Omega_{\mathrm{k}}(z) := \Omega_{\mathrm{k,0}}(1+z)^{2}$ with $\Omega_{\mathrm{k,0}} = -k/H^{2}_{0}$, being $k$ the parameter that characterizes the spatial curvature of the spacetime, $k=+1,-1,0$ for closed, open and flat Universe, respectively. If the interaction between dark matter and dark energy is allowed, both energy densities must satisfy the following conservation equations
\begin{align}
    & \rho'_{\mathrm{m}} - 3\left(\frac{\rho_{\mathrm{m}}}{1+z} \right) = - \frac{Q}{H_{0}E(z)(1+z)}, \label{eq:rhom}\\
    & \rho'_{\mathrm{de}} - 3\left[\frac{\rho_{\mathrm{de}}(1+\omega_{\mathrm{de}})}{1+z} \right] = \frac{Q}{H_{0}E(z)(1+z)}.\label{eq:rhode} 
\end{align}
The prime denotes derivatives w.r.t. the cosmological redshift and we have considered a barotropic equation of state, i.e., $p_{i} = \omega_{i}\rho_{i}$; notice that we focus on the case of cold dark matter, $\omega_{\mathrm{m}} = 0$. As can be seen from the above equations, the energy density of each component is not conserved, however the sum of both it is. The positivity (negativity) of the $Q$-term appearing on the r.h.s. of each equation determines the nature of the interaction between the components of the dark sector. We will give more details about this feature later. From the previous equations, we can determine the behavior of the parameter state for dark energy; using (\ref{eq:fried}), (\ref{eq:rhom}) and (\ref{eq:rhode}) we can compute
\begin{equation}
    1+\frac{\omega_{\mathrm{de}}(z)}{1+r(z)} = \frac{2}{3}\left[\frac{1}{2}(1+z)(\ln E^{2}(z))'-\Omega_{\mathrm{k}}(0)\left( \frac{1+z}{E(z)}\right)^{2}\right]\left[1-\Omega_{\mathrm{k}}(0)\left(\frac{1+z}{E(z)}\right)^{2}\right]^{-1},
    \label{eq:state}
\end{equation}
where we have adopted the standard definition for the coincidence parameter namely, $r(z)$, given as the quotient between the energy densities of the dark sector, $r(z):=\rho_{\mathrm{m}}(z)/\rho_{\mathrm{de}}(z)$. The appearance of the coincidence parameter in interacting models is typical; within the interacting approach, in order to solve the cosmological coincidence problem several proposals for the form of $r(z)$ can be found in the literature, see for instance \cite{coincidence}. Besides, we can see that the present time value of the parameter state for dark energy will depend only on the constant values $r(0)$ and $\Omega_{\mathrm{k}}(0)$; thus, the value of the spatial curvature plays an important role on the behavior of the dark energy under this construction.   

\subsection{Holographic approach}
\label{sec:cuts}
In this section we adopt the holographic description for dark energy, in other words, its corresponding energy density is written in terms of a characteristic length, usually termed as $L$ and $\rho_{\mathrm{de}} \propto L^{-2}$. We will restrict ourselves to the following cases discussed below.

\subsubsection{Model I}
\begin{figure}[htbp!]
\centering
\includegraphics[scale=0.75]{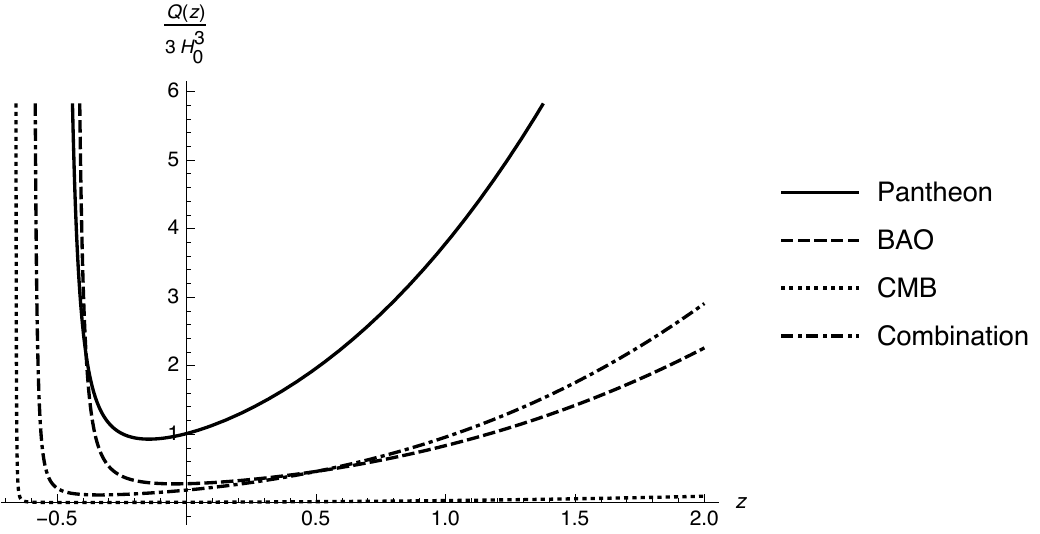}
\caption{Interaction term obtained from Model I with best fit values given in appendix \ref{sec:stat}. The combination plot corresponds to the best fit obtained using the joint data Pantheon+BAO+CMB.}
\label{fig:qtermc}
\end{figure}

As first holographic model we will consider the proposal explored in Ref. \cite{cruz18} by some of the authors. Such construction takes into account as characteristic length the usual Hubble scale, i.e., $L = 1/H$, then we have
\begin{equation}
    \rho_{\mathrm{de}}(z) = 3c^{2}H^{2}_{0}E^{2}(z),
    \label{eq:hde}
\end{equation}
where $c$ is a constant parameter. The equation (\ref{eq:hde}) corresponds to a very common formula for holographic dark energy, in order to describe an expanding universe we must have $0 < c^{2} < 1$, this kind of holographic model was initially proposed by Li in Ref. \cite{li} and it is interesting since reproduces the amount of dark energy in the Universe, as mentioned before. An advantage of this description is that by inserting the Eq. (\ref{eq:hde}) in the Friedmann constraint (\ref{eq:fried}) we can construct the energy density associated to dark matter, which results as 
\begin{equation}
\rho_{\mathrm{m}}(z) = 3H^{2}_{0}E^{2}(z)\left[1-c^{2}-\Omega_{\mathrm{k}}(0)\left(\frac{1+z}{E(z)}\right)^{2}\right].
\label{eq:matter}
\end{equation}
Before proceeding further in our approach, we must emphasize that in the standard treatment of an interacting model one gives priorly a convenient Ansatz for the interaction term, $Q$, then the energy densities $\rho_{\mathrm{m}}$ and $\rho_{\mathrm{de}}$ can be extracted from Eqs. (\ref{eq:rhom}) and (\ref{eq:rhode}). Other possibility is given by the following line of reasoning, due to the interaction the densities $\rho_{\mathrm{m}}$ and $\rho_{\mathrm{de}}$, exhibit deviations from the usual behavior, i.e., $\rho_{\mathrm{m}} \propto (1+z)^{3}$ and $\rho_{\mathrm{de}} \propto (1+z)^{3(1+\omega_{\mathrm{de}})}$, therefore if we provide some adequate energy densities, the interaction term can be reconstructed straightforwardly. However, note that in both alternatives we must introduce by hand some key ingredients, as example see Ref. \cite{coincidence, alternative}. In what follows we will proceed in a different way, we take advantage from the resulting energy density (\ref{eq:matter}). Inserting (\ref{eq:matter}) in (\ref{eq:rhom}) allow us to write
\begin{equation}
    (1+z)(\ln E^{2}(z))' = 3-\frac{1}{1-c^{2}}\left[\Omega_{\mathrm{k}}(0)\left(\frac{1+z}{E(z)}\right)^{2}+\frac{Q}{3H^{3}_{0}E^{3}(z)}\right],
\end{equation}
thus from Eq. (\ref{eq:state}) we can write the interaction term as follows
\begin{equation}
    \frac{Q(z)}{9(1-c^{2})H^{3}_{0}E^{3}(z)} = 1-\Omega_{\mathrm{k}}(0)\left(\frac{1+z}{E(z)}\right)^{2}\left[\frac{3-2c^{2}}{3(1-c^{2})}\right]-\left(1+\frac{\omega_{\mathrm{de}}(z)}{1+r(z)}\right)\left[1-\Omega_{\mathrm{k}}(0)\left(\frac{1+z}{E(z)}\right)^{2}\right].
\end{equation}
Note that in order to obtain the above interaction term, we only depend on the assumption of the form for the holographic dark energy density. Additionally, it is possible to find the normalized Hubble parameter as a function of the coincidence parameter. If we consider the quotient of the densities (\ref{eq:hde}) and (\ref{eq:matter}), we obtain
\begin{equation}
    r(z) = \frac{1}{c^{2}}\left[1-c^{2}-\Omega_{\mathrm{k}}(0)\left(\frac{1+z}{E(z)}\right)^{2}\right],
\end{equation}
therefore the normalized Hubble parameter can be penned as
\begin{equation}
    E^{2}(z) = -\frac{\Omega_{\mathrm{k}}(0)(1+z)^{2}}{c^{2}(r(z)-r_{c})}, \ \ \ \mbox{where} \ \ \ r_{c} := \frac{1-c^{2}}{c^{2}},
    \label{eq:normsing}
\end{equation}
note that the normalized Hubble parameter diverges at $r(z_{s}) = r_{c}$, i.e., for a future value $z_{s}$ of the cosmological redshift; the properties of this singularity were discussed in \cite{cruz18}, the model admits a phantom scenario and the flat case must be excluded from this description. As given in Eq. (\ref{eq:normsing}), we need to consider a proper parametrization for $r(z)$ in order to deduce the cosmic evolution of this cosmological model. In Ref. \cite{cruz18} was considered a Chevallier-Polarski-Linder (CPL)-like parametrization for the coincidence parameter given as
\begin{equation}
    r(z) = r(0) + \epsilon_{0}\frac{z}{1+z},
    \label{eq:cpl}
\end{equation}
this kind of parametrization has been widely used in the literature and has become standard. It describes a smooth evolution with a constant bounded value at early times for the coincidence parameter towards a present day value of $r(0)$, so we can safely describe the entire evolution of the universe from $z=\infty$ to the present day at $z=0$. Besides, given that our proposal admits a phantom scenario at $z_{s}$, then the parametrization (\ref{eq:cpl}) it is appropriate to describe the cosmic evolution since, $-1 < z_{s} < 0$. Using the previous equation we can solve the singularity condition for (\ref{eq:normsing}), $r(z_{s}) = r_{c}$, we obtain
\begin{equation}
    z_{s} = - \frac{r(0)-r_{c}}{\epsilon_{0}\left[1+\frac{r(0)-r_{c}}{\epsilon_{0}}\right]}. \label{eq:zsing}
\end{equation}
Now, from Eqs. (\ref{eq:cpl}) and (\ref{eq:zsing}) the normalized Hubble parameter (\ref{eq:normsing}) takes the form
\begin{equation}
    E^{2}(z) = -\Omega_{k}(0)\frac{(1+z)^{3}}{z-z_{s}}\eta, \ \ \ \mbox{with the constant}, \ \eta :=\frac{(1+z_{s})}{c^{2}\epsilon_{0}},
    \label{eq:normsing2}
\end{equation}
By defining the function $\theta(z) := (1+z)/(z-z_{s})$ and using the quantities written previously, the interaction term acquires the form
\begin{equation}
    \frac{Q(z)}{3H^{3}_{0}} = -\Omega_{k}(0)\sqrt{-\Omega_{k}(0)\eta \theta(z)}(1+z)^{3}[1 + \eta \theta^{2}(z)(1-c^{2})],
    \label{eq:qsing}
\end{equation}
where we have considered the following expression from Eq. (\ref{eq:state})
\begin{equation}
    1+\frac{\omega_{\mathrm{de}}(z)}{1+r(z)} = \frac{2-\eta \theta(z)(\theta(z)-3)}{3[1+\eta \theta(z)]}.
    \label{eq:omegateta}
\end{equation}
Note that the singular behavior of the phantom cosmology is inherited by the interaction term through the $\theta(z)$ function. The form given in Eq. (\ref{eq:qsing}) for the interaction term was also studied in Ref. \cite{cruz19}. An interesting feature of the $Q$-term given in Eq. (\ref{eq:qsing}) is the fact that can be written alternatively as the following product
\begin{equation}
    Q(z) = \Delta(z)\times \rho_{\mathrm{m}}(z)\times \rho_{\mathrm{de}}(z),
\end{equation}
where $\Delta(z)$ depends on the singular function $\theta(z)$, then the set of Eqs. (\ref{eq:rhom}) and (\ref{eq:rhode}) can be considered as a Lotka-Volterra like system, which describes a cyclic interchange of energy between the components of the dark sector, see \cite{cruz19} for details. Within the Ansatz philosophy for the interaction term it is common to find, $Q(z) \propto \rho_{\mathrm{m}}$, $Q(z) \propto \rho_{\mathrm{m}} + \rho_{\mathrm{de}}$, and some other possible combinations; this is because it was found that the dependence on the energy densities of the dark sector in $Q$ seems to be useful to alleviate the coincidence problem \cite{qandcoinc}. However, in our construction such dependence emerges naturally from the model. In Fig. (\ref{fig:qtermc}) we show the behavior of the interaction $Q$-term given in Eq. (\ref{eq:qsing}) using the best fit values for the cosmological parameters obtained from the statistical analysis discussed in appendix \ref{sec:stat}. As can be seen, the interaction term exhibits a future divergence and keeps positive along the cosmic evolution.

\subsubsection{Model II}
\begin{figure}[htbp!]
\centering
\includegraphics[scale=0.75]{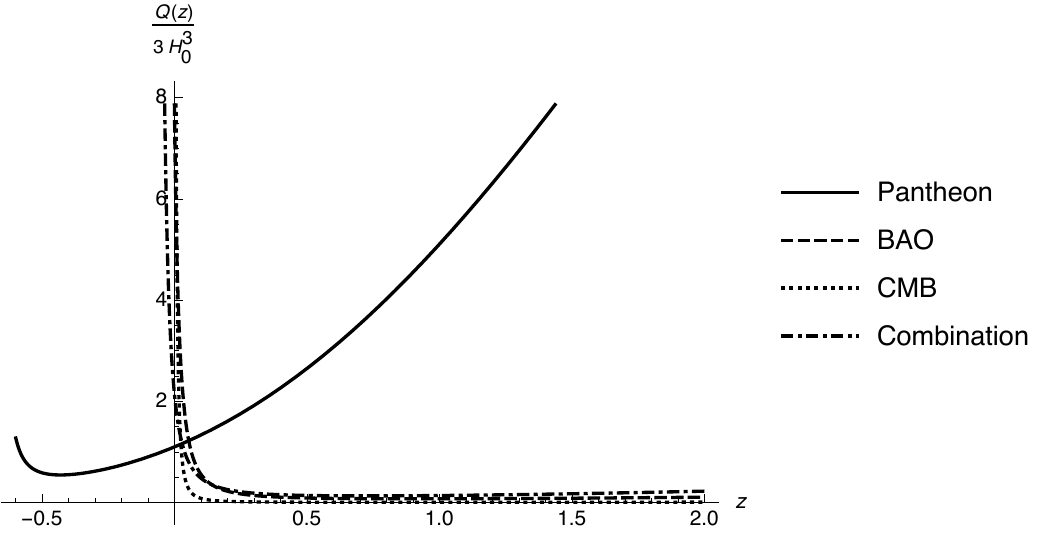}
\caption{Interaction term obtained from the holographic Model II with a generalized parameter $c(z)$.}
\label{fig:qtermvariablec}
\end{figure}

The second holographic cut for the dark energy density will be given by the following expression
\begin{equation}
    \rho_{\mathrm{de}}(z) = 3[\beta_{1}-\beta_{2}\Omega_{\mathrm{k}}(0)(1+z)^{2}]H^{2}_{0}E^{2}(z),
    \label{eq:apparent}
\end{equation}
this holographic model was proposed in \cite{holo18} for non interacting fluids and resembles the expression of the apparent horizon; $\beta_{1,2}$ are constant parameters. If we compare the Eqs. (\ref{eq:hde}) and (\ref{eq:apparent}), we can observe that Model II can be written in the form of the usual formula for the holographic dark energy but in this case, $c^{2} \rightarrow c^{2}(z) := \beta_{1}-\beta_{2}\Omega_{\mathrm{k}}(0)(1+z)^{2}$, i.e., the concomitant coefficient to the normalized Hubble parameter is now a function of the cosmological redshift. In Ref. \cite{radicella} was proposed that a more realistic description of the Universe from the holographic perspective could be obtained if one generalizes the holographic formula (see Eq. (\ref{eq:hde})) from a constant parameter $c^{2}$ to a varying function of the cosmological redshift (or time). As discussed in Ref. \cite{holo18}, the energy density (\ref{eq:apparent}) can be written in terms of only one free parameter, say $\beta_{1}$. If we evaluate $c^{2}(z)$ at present time one gets, $c^{2}(0) = \beta_{1}-\beta_{2}\Omega_{\mathrm{k}}(0)$. Therefore, (\ref{eq:apparent}) is re-expressed as
\begin{equation}
    \rho_{\mathrm{de}}(z) = 3[\beta_{1}+(c^{2}(0)-\beta_{1})(1+z)^{2}]H^{2}_{0}E^{2}(z),
    \label{eq:apparent2}
\end{equation}
for simplicity in the notation from now on we will write $c^{2}$ instead $c^{2}(0)$. By repeating the procedure discussed for Model I, we can write the interaction $Q$-term as follows
\begin{eqnarray}
    \frac{Q(z)}{9(1-c^2(z))H^{3}_{0}E^3(z)} &=& 1 - \Omega_{\mathrm{k}}(0)\left(\frac{1+z}{E(z)}\right)^{2}\left(\frac{3-2c^2(z)}{3(1-c^2(z))}\right) - \left(1+\frac{\omega_{\mathrm{de}}(z)}{1+r(z)}\right)\times \nonumber \\ &\times& \left[1-\Omega_{\mathrm{k}}(0)\left(\frac{1+z}{E(z)}\right)^{2}\right] + c^2(z)\left(\frac{1+z}{3(1-c^2(z))}\right)\frac{d \ln{c^2(z)}}{dz},
    \label{eq:qterm2}
\end{eqnarray}
the normalized Hubble parameter for Model II is a generalization of the one obtained for Model I given in Eq. (\ref{eq:normsing}), we obtain
\begin{equation}
    E^{2}(z) = -\frac{\Omega_{\mathrm{k}}(0)(1+z)^{2}}{c^{2}(z)(r(z)-r_{c}(z))}, \ \ \ \mbox{being} \ \ \ r_{c}(z) := \frac{1-c^{2}(z)}{c^{2}(z)},
    \label{eq:normsing3}
\end{equation}
as can be seen from the above expression, for a finite value of the redshift, $z_{s}$, we have a divergent behavior for $E(z)$. From the condition, $r(z_{s}) - r_{c}(z_{s}) = 0$, and the CPL-like parametrization (\ref{eq:cpl}) for $r(z)$, we can obtain a cubic equation for $z_{s}$, which has the form $(r(0)-1)(c^{2}-\beta_{1})(1+z_{s})^{3}+\epsilon_{0}z_{s}(c^{2}-\beta_{1})(1+z_{s})^{2}+[\beta_{1}(r(0)-1)-1](1+z_{s})+\epsilon_{0}\beta_{1}z_{s}=0$. Such equation is solved by considering the best fit values of the parameters obtained from the statistical analysis and has only one real root. Note that in this case the value $r(0)$ can be obtained from the evaluation of the normalized Hubble parameter (\ref{eq:normsing3}) at present time, yielding
 \begin{equation}
    r(0) = -1 + \frac{1-\Omega_{\mathrm{k}}(0)}{c^{2}}.
    \label{eq:coinci2to}
\end{equation}
Finally, as expected, the interaction $Q$-term is also divergent since depends on the normalized Hubble parameter, similarly as done for Model I, with the use of expression (\ref{eq:state}) the Eq. (\ref{eq:qterm2}) takes the following form after some lengthy but straightforward manipulations
\begin{equation}
    Q(z) = 3H^{3}_{0}E^{3}(z)\left[ 1-c^2(z) + 2(1-\beta_1) - \Omega_{k}(0) \left(\frac{1+z}{E(z)}\right)^2 - (1+z)(1-c^2(z))(\ln E^{2}(z))'\right],
    \label{eq:Qfinal}
\end{equation}
for simplicity in the notation we do not write explicitly the derivative of the normalized Hubble parameter appearing in the above equation. Regarding the above interaction term, we would like to comment that it is possible to write it in terms of the energy densities of the dark sector as done with Model I. In this case we obtain, 
\begin{equation}
\frac{Q(z)}{H_{0}E(z)} = \rho_{\mathrm{m}}(z) + \Delta(z)\rho_{\mathrm{de}}(z),  
\end{equation}
where we have defined, $\Delta(z) := 2H_{0}(1+z)[E'(z)(1-1/c^{2}(z))+(1-\beta_{1})/\lbrace(1+z)c^{2}(z)\rbrace]$. Despite the cumbersome form of the $\Delta$ function, its explicit dependence on $z$ can be exploited since $E(z)$ and $c^{2}(z)$ are known. In Fig. (\ref{fig:qtermvariablec}) the interaction $Q$-term for Model II is shown. Note that as in the Model I, this interaction term is positive. Using the best fit values for the cosmological parameters it is found that Model II exhibits a future divergence at $z_{s} = -0.6764$, $z_{s} = -0.0405$, $z_{s} = -0.0086$ and $z_{s} = -0.0818$ for Pantheon, BAO, CMB and Pantheon+BAO+CMB data sets respectively. It is worthy to mention that for both holographic models the interaction terms are positive, this means that no phase transitions are expected for this kind of interacting model, in general, changes in the sign of the $Q$-term induce variations in the value of heat capacities of the cosmic components, see for instance \cite{heat}.

\section{Thermodynamics of the interacting scenario}
\label{sec:thermo}
\begin{figure}[htbp!]
\centering
\includegraphics[scale=0.78]{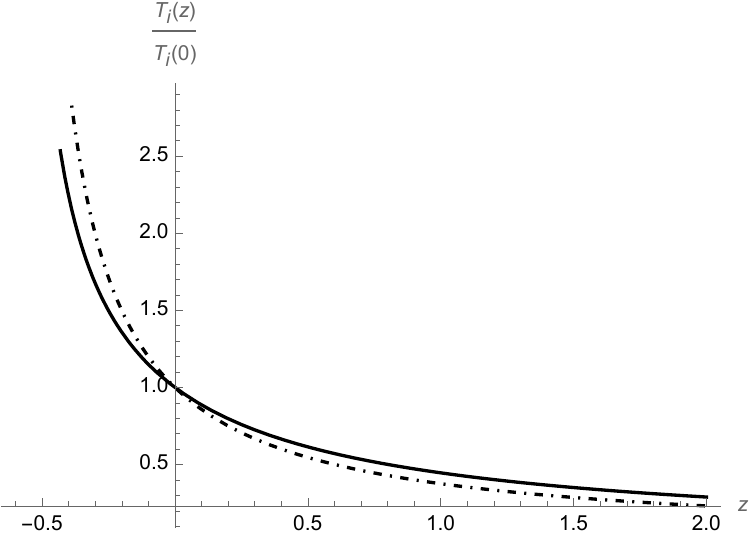}
\caption{Temperatures for dark matter and dark energy obtained from Model I. The solid line represents the dark matter temperature and the dot-dashed line stands for dark energy sector temperature.}
\label{fig:tempsconstantc}
\end{figure}

In this section we show some thermodynamics results for both holographic models within the interacting scenario. In standard cosmology, for a perfect fluid we have the following temperature evolution equation
\begin{equation}
\frac{\dot{T}}{T} = -3H\left(\frac{\partial p}{\partial \rho}\right)_{n}.
\label{eq:evolution}
\end{equation}
The previous equation is valid always that the Gibbs integrability condition holds together with the number ($n$) and energy conservation. However, as can be seen the constant temperature approach of standard thermodynamics is no longer available, i.e., the global equilibrium condition, $\dot{T} = 0$, is not satisfied \cite{maartens}. If we consider a barotropic equation of state in the temperature evolution (\ref{eq:evolution}) we can write
\begin{equation}
\int d \ln T = 3 \int \omega(z) \frac{dz}{1+z} \ \Rightarrow \ T(z) = T(0)\exp \left(3 \int \omega(z) \frac{dz}{1+z} \right),
\label{eq:tempe}
\end{equation}
where $T(0)$ is an integration constant. Following the line of reasoning of Ref. \cite{victor}, we adopt the effective temperatures method; the expressions (\ref{eq:rhom}) and (\ref{eq:rhode}) can be written conveniently as follows 
\begin{align}
& \rho'_{\mathrm{m}} - 3 \left(\frac{1+\omega_{\mathrm{eff,m}}}{1+z} \right)\rho_{\mathrm{m}} = 0, \label{eq:q4} \\
& \rho'_{\mathrm{de}} - 3 \left(\frac{1+\omega_{\mathrm{eff,de}}}{1+z} \right)\rho_{\mathrm{de}} = 0, \label{eq:q3}
\end{align} 
where the effective parameters are given as
\begin{align}
& \omega_{\mathrm{eff,m}} = -\frac{Q}{3\rho_{\mathrm{m}}H_{0}E(z)}. \label{eq:eff2}\\
& \omega_{\mathrm{eff,de}} = \omega_{\mathrm{de}} + \frac{Q}{3\rho_{\mathrm{de}}H_{0}E(z)}. \label{eq:eff1}
\end{align} 
Therefore, using the Eq. (\ref{eq:tempe}) we can calculate the temperature for the dark energy and dark matter components, yielding
\begin{align}
& T_{\mathrm{m}}(z)  = T_{\mathrm{m}}(0)\exp \left[ \int^{z}_{0} \left(-\frac{Q}{\rho_{\mathrm{m}}H_{0}E(x)} \right) \frac{dx}{1+x} \right], \label{eq:tempm}\\
& T_{\mathrm{de}}(z) = T_{\mathrm{de}}(0)\exp \left[3 \int^{z}_{0} \left(\omega_{\mathrm{de}} + \frac{Q}{3\rho_{\mathrm{de}}H_{0}E(x)} \right) \frac{dx}{1+x} \right]. \label{eq:tempx}
\end{align}
As can be seen from equation (\ref{eq:state}), the parameter state depends on the cosmological redshift, for Model I its explicit expression is given in Eq. (\ref{eq:omegateta}). An advantage we can use for Model I is that some cosmological quantities such as the parameter state for dark energy and the interaction term depend on the $\theta(z)$ function, therefore we adopt the following change of variable to integrate, $z \rightarrow \theta(z)$; then from Eqs. (\ref{eq:tempm}) and (\ref{eq:tempx}) we can write
\begin{align}
& T_{\mathrm{m}}(z)  = T_{\mathrm{m}}(0)\exp \left[\int^{\theta(z)}_{\theta_{0}} \left(-\frac{Q}{\rho_{\mathrm{m}}H_{0}E(\theta)} \right) \frac{d\theta}{\theta(1-\theta)} \right], \label{eq:tempmteta}\\
& T_{\mathrm{de}}(z) = T_{\mathrm{de}}(0)\exp \left[3 \int^{\theta(z)}_{\theta_{0}} \left(\omega_{\mathrm{de}}(\theta) + \frac{Q}{3\rho_{\mathrm{de}}H_{0}E(\theta)} \right) \frac{d\theta}{\theta(1-\theta)} \right], \label{eq:tempxteta}
\end{align}
where $\theta_{0}$ means evaluation at present time, i.e., $\theta_{0} := \theta(z=0)$. Using the Eqs. (\ref{eq:normsing2}), (\ref{eq:qsing}) and (\ref{eq:omegateta}) in the expressions (\ref{eq:tempmteta}) and (\ref{eq:tempxteta}), one gets
\begin{align}
& T_{\mathrm{m}}(z)  = T_{\mathrm{m}}(0)\left(\frac{\theta_{0}}{\theta(z)}\right)\left\lbrace \frac{1-\theta(z)}{1-\theta_{0}}\frac{[1+\eta(1-c^{2})\theta(z)]}{[1+\eta(1-c^{2})\theta_{0}]} \right\rbrace, \label{eq:tempmlast}\\
& T_{\mathrm{de}}(z) = T_{\mathrm{de}}(0)\left(\frac{1-\theta(z)}{1-\theta_{0}}\right). \label{eq:tempxlast}
\end{align}
In Fig. (\ref{fig:tempsconstantc}) we show the behavior of the quotient $T_{\mathrm{i}}(z)/T_{\mathrm{i}}(0)$ for each component given in Eqs. (\ref{eq:tempmlast}) and (\ref{eq:tempxlast}) by considering the constrained values for the parameters of the model obtained from BAO, we consider this case since the value obtained for $\Omega_{\mathrm{k,0}}$ is closer to the upper bound imposed by Planck 2018 results, as commented below. As shown in the plots, both temperatures are positive and exhibit a growing behavior. We could consider this result as atypical for the dark matter sector temperature. However, this nature is due to the existence of interaction with the dark energy component. For $Q=0$ we recover the single fluid result, $T_{\mathrm{m}}(z) = T_{\mathrm{m}}(0) = \mbox{constant}$, see Eq. (\ref{eq:tempm}). This is in agreement with other works where the interaction in the dark sector is turned on \cite{victor, mimetic}. Given the singular nature of the cosmological model both temperatures diverge at $z_{s}$. However, at the past the dark matter temperature is slightly higher than dark energy temperature but for a future evolution before the singularity takes place (not shown in the plots) the situation is inverted, the dark energy temperature is higher than dark matter temperature; having a higher temperature for dark energy component than dark matter one becomes relevant if we demand the fulfillment of the second law of thermodynamics, i.e, a positive growth of the entropy, $S$, we will discuss this below briefly. As far as we know, this thermodynamics requirement is also guaranteed for the Universe at large scales \cite{manuel}. It is worthy to mention that the behavior exhibited by the temperatures could change in dependence of the choice for their initial values; we must have in mind the following conditions, $T_{\mathrm{m}}(0) = T_{\mathrm{de}}(0)$, we can also consider, $T_{\mathrm{m}}(0) > T_{\mathrm{de}}(0)$ and $T_{\mathrm{m}}(0) < T_{\mathrm{de}}(0)$, since at this moment no definite results coming from cosmological observations with respect to these initial temperatures can be found. 

\begin{figure}[htbp!]
\centering
\includegraphics[scale=0.78]{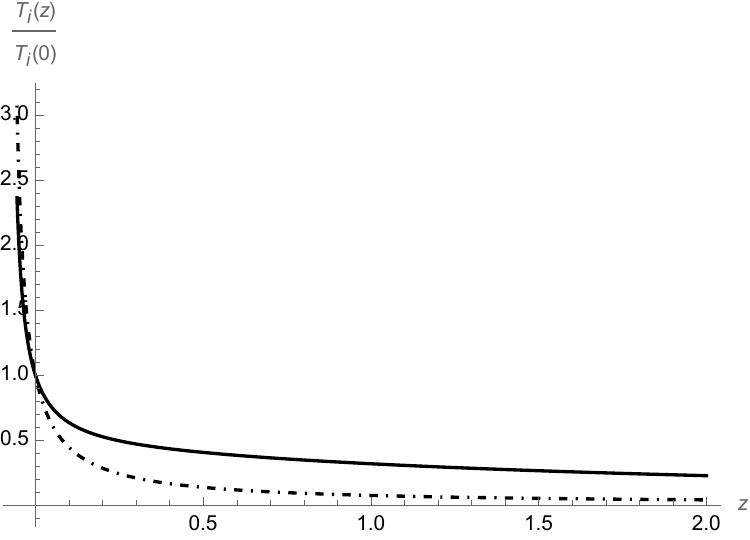}
\caption{Temperatures for dark matter and dark energy of Model II using the best fit values obtained with Pantheon data set. The solid line represents $T_{\mathrm{m}}(z)/T_{\mathrm{m}}(0)$ and the dot-dashed line $T_{\mathrm{de}}(z)/T_{\mathrm{de}}(0)$.}
\label{fig:tempsvariablec}
\end{figure}

For Model II we can also implement the previous procedure but in this case we do not use the change of variable, $z \rightarrow \theta(z)$, in Eqs. (\ref{eq:tempm}) and (\ref{eq:tempx}). By constructing the parameter state for dark energy from Eq. (\ref{eq:state}) and using (\ref{eq:Qfinal}) together with the energy densities for the dark sector we can obtain
\begin{align}
    T_{\mathrm{m}}(z) & = \frac{T_{\mathrm{m}}(0)}{1+z}\left[ \frac{\alpha_1^2+(1+z)^2}{\alpha_1^2 + 1}\right]^{\frac{\alpha_1^2 \alpha_2(1-\beta_1)}{\beta_1 r_0 (\alpha_1^2 + (\alpha_2-1)^2)}}\left(\frac{\alpha_2}{\alpha_2+z} \right)^{\frac{2\alpha_1^2 \alpha_2(1-\beta_1)}{\beta_1 r_0 (\alpha_1^2 + (\alpha_2-1)^2)}}\times \nonumber \\
    & \times \exp \left\{\int_{0}^{z} \frac{ E^2(x) }{ E^2(x) - f(x) } \frac{d \ln E^{2}(x)}{d x}  dx -  \frac{2\alpha_1 \alpha_2 (\alpha_2 - 1)(1-\beta_1)}{\beta_1 r_0 [\alpha_1^2 + (\alpha_2-1)^2]}\Theta(z)  \right\}, \label{eq:temvar1}\\ 
    T_{\mathrm{de}}(z) & = \frac{T_{\mathrm{de}}(0)}{(1+z)^3}\left( \frac{c(z)}{c} \right)^2 E^2(z),\label{eq:temvar2}
\end{align}
where we have defined the following quantities for simplicity, $\alpha_1^2 = \frac{\beta_1}{c^2-\beta_1}$ and $\alpha_2 = \frac{r(0)}{r(0)+\epsilon_{0}}$, and the functions
\begin{eqnarray}
    f(z) &:=& \frac{\Omega_\mathrm{k}(z)}{1-c^2(z)},\\
    \Theta(z) &:=& \tan^{-1}{\left(\frac{1+z}{\alpha_1}\right)} - \tan^{-1}{\left(\frac{1}{\alpha_1}\right)}.
\end{eqnarray}
The behavior obtained for the temperatures (\ref{eq:temvar1}) and (\ref{eq:temvar2}) is shown in Fig. (\ref{fig:tempsvariablec}), the positivity of both temperatures is kept along the cosmic expansion. Similarly to Model I, at the past the value for dark matter temperature is higher than the dark energy one, but again, for a future evolution of the cosmic expansion the dark energy temperature becomes higher than dark matter one. In agreement with the singular nature exhibited by the interaction $Q$-term (\ref{eq:Qfinal}), both temperatures diverge around $z \approx -0.04$ given that we are considering the constrained values for the cosmological parameters with BAO data, we do not show their singular nature in the plot.    

In terms of the redshift, the condition $dS/dt > 0$ takes the form $dS/dz < 0$. Then, as discussed in Ref. \cite{victor}, from the second law we have
\begin{equation}
    TdS = d(\rho V) + pdV,
\end{equation}
being $V$ the Hubble volume defined as $V(z)=V_{0}(a/a_{0})^{3}=V_{0}(1+z)^{-3}$, in this case the quantity $\rho V$ denotes the internal energy. By means of a barotropic equation of state and Eqs. (\ref{eq:rhom}), (\ref{eq:rhode}) in the above equation, we can write for each component of the dark sector
\begin{equation}
    -\frac{T_{\mathrm{m}}}{V}\frac{dS_{\mathrm{m}}}{dz} = \frac{Q}{H_{0}E(z)(1+z)} = \frac{T_{\mathrm{de}}}{V}\frac{dS_{\mathrm{de}}}{dz},
    \label{eq:total}
\end{equation}
then the entropy associated to each component is not constant, the adiabatic ($S = \mbox{constant}$) cosmic evolution is recovered for null interaction, $Q=0$. From Eq. (\ref{eq:total}) one gets the following condition
\begin{equation}
    T_{\mathrm{m}}dS_{\mathrm{m}} + T_{\mathrm{de}}dS_{\mathrm{de}} = 0,
\end{equation}
which leads to
\begin{equation}
    \frac{d}{dz}(S_{\mathrm{m}}+S_{\mathrm{de}}) = -\left(-1+\frac{T_{\mathrm{de}}}{T_{\mathrm{m}}} \right)\frac{dS_{\mathrm{de}}}{dz} = -\left(-\frac{1}{T_{\mathrm{de}}}+\frac{1}{T_{\mathrm{m}}} \right)\frac{VQ}{H_{0}E(z)(1+z)}.
    \label{eq:entropy}
\end{equation}
Thus the above equation indicates that the second law is satisfied by the total entropy only if $T_{\mathrm{de}} > T_{\mathrm{m}}$ since the condition $Q > 0$ is satisfied by both holographic models. This implies that the cosmological expansion is not an adiabatic process, $S_{\mathrm{m}}+S_{\mathrm{de}} \neq \mbox{constant}$. From the thermodynamics point of view, this scenario seems to be more consistent than the $\Lambda$CDM model, where the cosmic evolution is described by an adiabatic process, i.e., constant entropy (reversible thermodynamics) and null temperature for dark energy. We can also observe that the second law of thermodynamics is not guaranteed at the past in this framework for both holographic models but its validity is fulfilled at the future cosmic evolution. To end this section we comment some possible differences between the single fluid description and these interacting scenarios. For a dark energy fluid with constant parameter state, $\omega$, we have
\begin{equation}
    \rho_{\mathrm{de}}(z) = \rho(0)(1+z)^{3(1+\omega)},
\end{equation}
then, from Eq. (\ref{eq:tempe}) we can obtain its temperature straightforwardly as follows 
\begin{equation}
    T(z) = T(0)(1+z)^{3\omega}.
\end{equation}
The quotient between the energy density for dark energy and its temperature takes the form
\begin{equation}
    \frac{\rho_{\mathrm{de}}(z)}{T(z)} = \frac{\rho(0)}{T(0)}(1+z)^{3},
\end{equation}
which is independent of the parameter state, $\omega$. At present time the above quotient is simply a constant value, $\rho(0)/T(0)$ and at the far future, $z\rightarrow -1$, this quotient tends to zero. If we consider the holographic cut (\ref{eq:hde}) of Model I for dark energy and the temperature (\ref{eq:tempxlast}), we can observe that in this scenario the quotient between the energy density and temperature tends to zero as $z\rightarrow -1$ and takes the constant value at present time. On the other hand, if we construct the corresponding quotient to Model II by means of Eqs. (\ref{eq:apparent2}) and (\ref{eq:temvar2}), we have, $(3c^{2}H^{2}_{0}/T_{\mathrm{de}}(0))(1+z)^{3}$. Therefore, for this specific quotient, both holographic models lead to a constant value at present time as in the single fluid description.       

\section{Concluding remarks}
\label{sec:final}
In this work we explored the cosmological implications of two holographic descriptions for dark energy within the interacting scenario with the inclusion of spatial curvature. One scenario (labeled as Model I) was proposed by some of the authors earlier in order to understand the role of the spatial curvature in the interacting scheme. Additionally, by turning on the interaction in the dark sector in a previously proposed model (Model II), we obtain a generalized holographic model inspired on the apparent horizon, i.e., spatial curvature dependent. In both models we consider the usual holographic formula and the characteristic length is given by the Hubble scale. However, the apparent horizon like model can be thought as a simple generalization of the usual model by considering a function $c(z)$ instead a constant parameter $c$. We are aware that the results presented here rely strongly on the election of $L$ and could be improved (or worsened) for different choices of the characteristic length. However, our approach pretends to figure out the importance of spatial curvature within the interacting scheme with the most simple description for holographic dark energy. This issue deserves deeper investigation. An interesting feature of both models is that no Ansatz is needed in order to write explicitly the interaction term $Q$ as usually done in the literature, this term can be constructed in both cases due to the presence of the spatial curvature and by the consideration of the CPL like parametrization for the coincidence parameter. Other relevant feature of this setup is the presence of a future singularity in both models, leading to an over accelerated cosmic expansion known as phantom scenario, such stage is not discarded at all by recent observations.\\  

In our study we considered recent cosmological data sets merely as support for the physical consequences of both models, it is not our intention to promote these scenarios as falsifying models. In all cases a closed Universe is favored by observations as revealed by the statistical analysis for both holographic models, this is in agreement with other values reported for the curvature parameter. As discussed in Ref. \cite{hdetens}, the holographic dark energy approach could provide a good scenario to alleviate the $H_{0}$ tension. This fact depends on the presence of a {\it turning point} in the normalized Hubble parameter, in other words, a decreasing behavior for $E(z)$ can suddenly change its tendency at some stage of the cosmic evolution, then we could have a model that shows a similar behavior to the $\Lambda$CDM model but at some epoch deviates from this; this characteristic for $E(z)$ can be the responsible of inducing increments in the value of $H_{0}$. In fact, the value for $H_{0}$ could also change if the turning point exists at the past or in the future evolution, being more consistent the case of a turning point at the past close to $z=0$. We must mention that an interesting alternative to address in our scheme could be given by the election of other characteristic length for dark energy that can induce turning points in $E(z)$. We leave this as future investigation.\\

At thermodynamics level, the temperatures for the fluids of the dark sector were computed by means of the effective method. Using the best fit values for the cosmological parameters of Models I and II, it was found that temperatures are well defined and exhibit a growing behavior from the past to the future, this is in agreement with the interacting approach. Some comments are in order: for both models the temperatures for dark matter and dark energy can be distinguished easily and at the past the dark matter temperature is higher than dark energy one. For future cosmic evolution the dark energy temperature becomes higher than dark matter one. Given that in both scenarios the temperatures are of the same order of magnitude, we have a consistent description from the thermodynamics point of view since we could consider that such temperatures are very small with respect to the CMB temperature, making them hard to detect. Finally, given that we are in the context of interacting fluids, the cosmic expansion it is not adiabatic, i.e., the entropy has a varying behavior which seems to be more consistent at physical level than the $\Lambda$CDM model where the processes are reversible. We explored the condition for which the second law is fulfilled, as discussed previously the second law is achieved if $T_{\mathrm{de}}(z) > T_{\mathrm{m}}(z)$ and $Q(z) > 0$, as shown in the text, both holographic models obey the latter condition for the interaction term.\\

\begin{figure}[htbp!]
\centering
\includegraphics[scale=0.78]{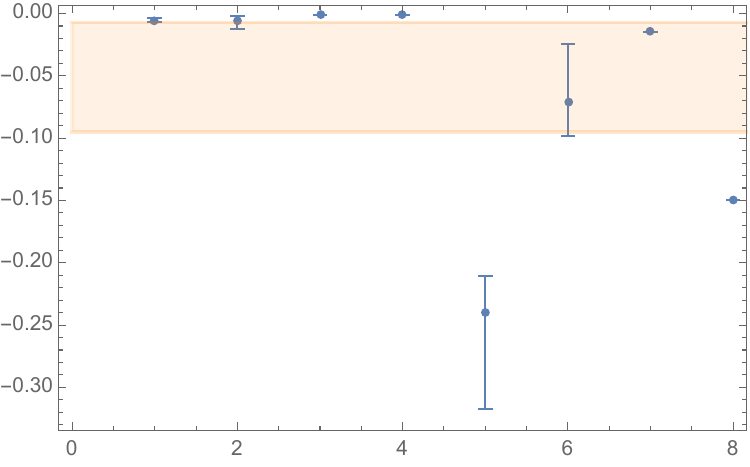}
\caption{Constrained values for the curvature density parameter, $\Omega_{\mathrm{k,0}}$, in Model I and Model II using Pantheon, BAO, CMB and joint data, respectively. The shaded region corresponds to the interval $-0.007 > \Omega_{\mathrm{k,0}} > -0.095$.}
\label{fig:values}
\end{figure}

Finally, as found in this work, the considered holographic scenarios seem to be viable since the results do not lack of physical meaning and the values obtained for some relevant cosmological parameters do not contradict those found in the literature.
According to the results obtained from the statistical analysis for both holographic models, we can observe that a negative curvature density parameter, $\Omega_{\mathrm{k,0}}$, is favored by the observations, given the definition of $\Omega_{\mathrm{k,0}}$ this implies a positive parameter $k$, which corresponds to a closed Universe. The latest released results by the Planck collaboration exhibit the presence of an enhanced lensing amplitude in the CMB power spectra, such effect could be explained by considering a closed Universe, i.e., positive curvature, $k$. The constraint on the curvature parameter imposed by the CMB spectra observations is $-0.007 > \Omega_{\mathrm{k,0}} > -0.095$; by comparing the 2015 and 2018 observations compiled by the Planck collaboration, we can conclude that the preference for a closed Universe increased in the latest results, the existence of inconsistencies and tensions between different sets of cosmological observations was properly quantified in \cite{valentino, curv2} with the inclusion of spatial curvature. In light of these inconsistencies, we can no longer conclude that observations favor a flat Universe. In Fig. (\ref{fig:values}) we summarize our results, which correspond to 8 different values for the curvature density parameter and are organized from left to right for Model I and Model II. As can be seen, all the values obtained with Model I for the curvature are higher than the upper bound imposed by the CMB spectra, then Model II is in more agreement with Planck 2018 results when BAO and CMB data are used. Then we are confident that a more robust description of the holographic dark energy than the one presented here could lead to better results. 

\section*{Acknowledgments}
M.C. work has been supported by S.N.I. (CONACyT-M\'exico).

\newpage

\appendix
\section{Statistical analysis}
\label{sec:stat}

In this appendix we show explicitly the analysis performed to restrict the cosmological parameters of both holographic models. For Model I Eq. (\ref{eq:normsing2}) can be used to implement a test with observations considering five free parameters: $h, c^{2}, \epsilon_{0}, \Omega_{\mathrm{k}}(0)$ and $z_s$; where $h$ is defined as usual $h:=H_{0}/100$; Eq. (\ref{eq:normsing3}) is used to test Model II. The aim of this statistical analysis is merely explore the cosmological implications of our scheme based on recent astrophysical observations. In fact, we are aware that if we perform a $\chi^{2}$ comparison of this approach with $\Lambda$CDM (or if we consider other tests as the BIC), possibly our cosmological model will not be able to compete due to the number of additional parameters.\\ 

\begin{figure}[htbp!]
    \centering
    \includegraphics[width=14cm]{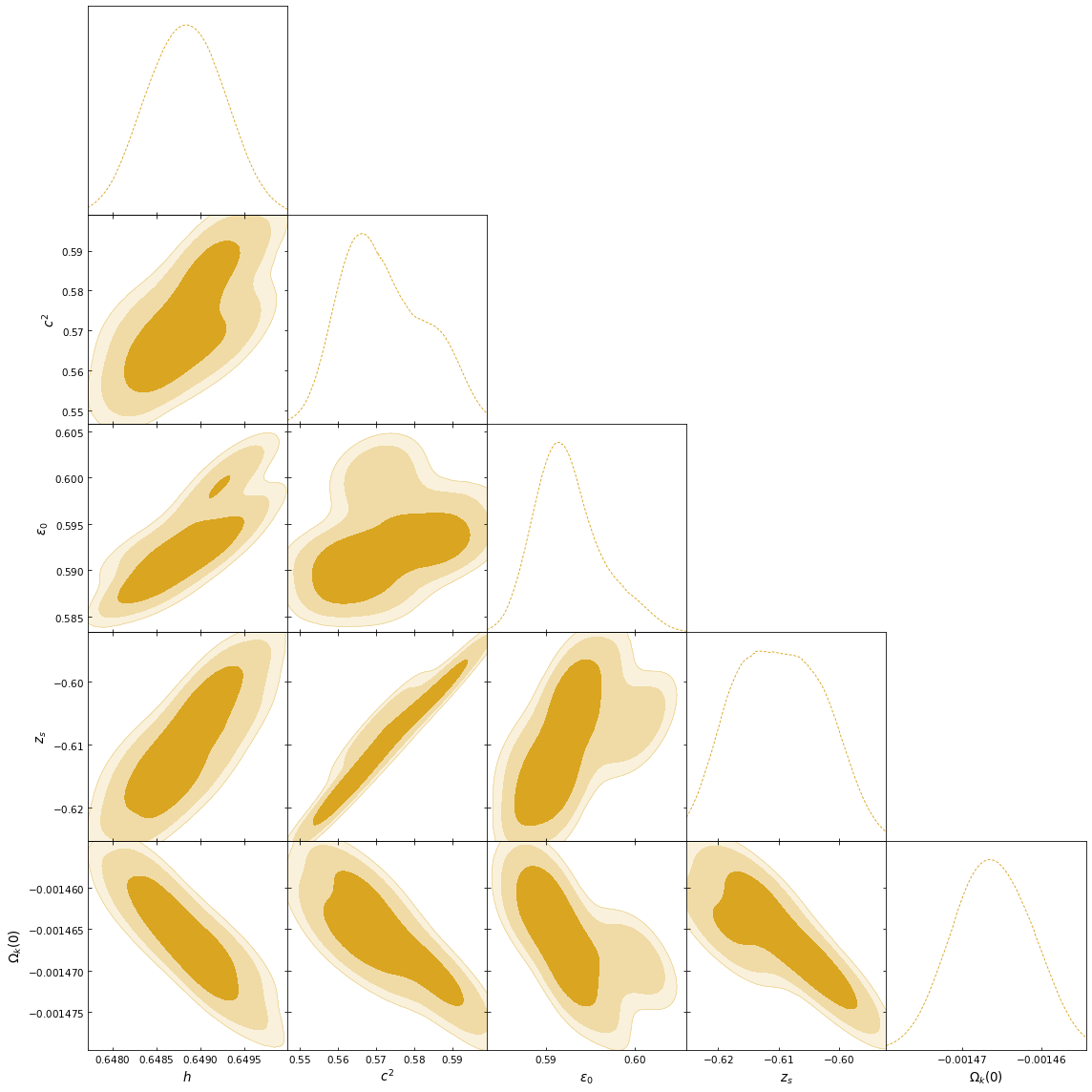}
    \caption{Confidence contours for $1\sigma$, $2\sigma$ and $3\sigma$ for Model I in parameter space using joint data.}
    \label{fig:constantc}
\end{figure}

\begin{figure}[htbp!]
    \centering
    \includegraphics[width=14cm]{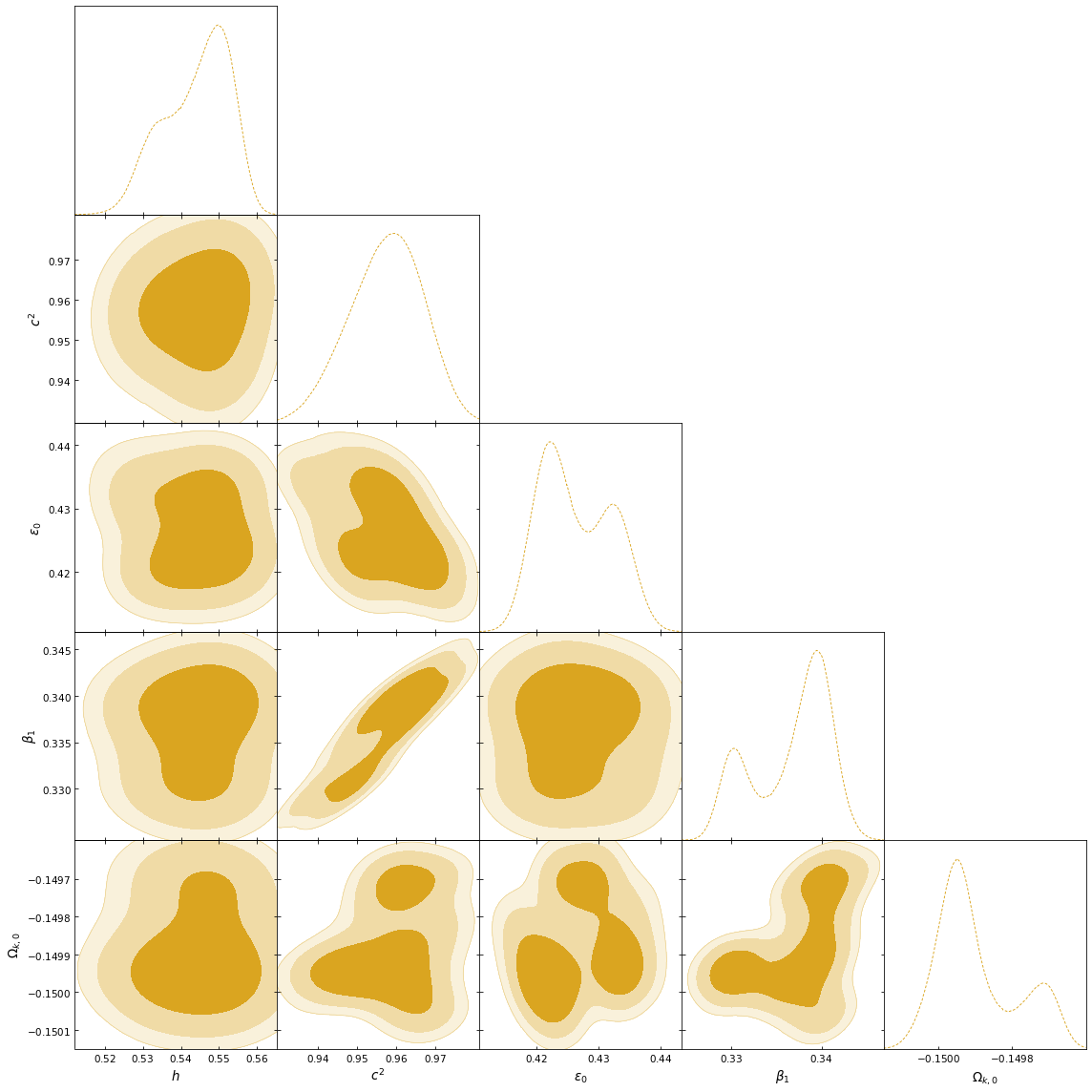}
    \caption{Confidence contours for $1\sigma$, $2\sigma$ and $3\sigma$ for Model II in parameter space using all data.}
    \label{fig:variablec}
\end{figure}

In order to perform this analysis we will use type Ia supernova (SNIa) measurements known as Pantheon sample consisting on a total of 1048 data points ranging from $0.01 < z < 2.3$ \cite{pantheon} and baryon acoustic oscillations (BAO) compiled from \cite{bao}. The analysis is carried out by implementing a Markov Chain Monte Carlo (MCMC) code called SimpleMC\footnote{More details about SimpleMC can be found at: \url{https://github.com/slosar/april}. The new version of the code is stored at \url{https://github.com/ja-vazquez/SimpleMC}.}, which is entirely written in Python, this code solves the cosmological equations for the background parameters in the same way as CLASS or CAMB and it contains the statistical parameter inference from CosmoMC or MontePython \cite{vazquez}. For our study we will also consider the compressed version of the recent Planck CMB data used by SimpleMC. We adopt the Gelman-Rubin convergence criteria for our chains. In general, the luminosity peak of the supernovas type Ia can be used as a distance indicator from the relation between redshift and distance, relative distance measurements can be performed using the luminosity distance which is written as follows
\begin{equation}
    d_{L} = \sqrt{\frac{L}{4\pi \mathcal{S}}},
\end{equation}
where $L$ is the luminosity and $\mathcal{S}$ is the radiation flux density. An important quantity for the observation of supernovas is the modular distance or distance modulus
\begin{equation}
    \mu = m_{B} - M_{B} + \alpha X - \beta C,
\end{equation}
where $m_{B}$ is the maximum apparent magnitude in the blue band (B), $\alpha$, $\beta$ and $M_{B}$ are parameters that depend on host galaxy properties. $X$ is related to the widening of the light curve and $C$ is a correction for the supernova color. The distance modulus and the luminosity are related through the following expression
\begin{equation}
    \mu = 5\log_{10}\left(\frac{d_{L}}{10 \ pc} \right).
\end{equation}
BAO is a statistical property coming from our primitive Universe. A viable description for the early Universe is given by the existence of a plasma formed of photons and matter, the interaction between its components lead to the formation of spherical waves. Once the Universe cooled down, the photons decoupled from baryons and the final configuration of such process is a over density of matter at the center of the spherical wave and a shell of baryons of fixed radius usually known as sound horizon, $r_{s}$. The BAO scale is estimated by the ratio
\begin{equation}
    \frac{D_{M}(z)}{r_{s}},
\end{equation}    
where $D_{M}(z)$ is the comoving angular diameter distance. For the transverse direction one can write in the line of sight
\begin{equation}
     \frac{D_{M}(z)}{r_{s}} = \frac{\mathcal{P}}{(1+z)\sqrt{-\Omega_{\mathrm{k}}}}\sin\left[\sqrt{-\Omega_{\mathrm{k}}}\int^{z}_{0}\frac{dz'}{E(z')} \right], 
     \label{eq:bao1}
\end{equation}
where $\mathcal{P} := c/(r_{s}H_{0})$, the Hubble parameter can be constrained by considering an analogous expression
\begin{equation}
    \frac{D_{H}(z)}{r_{s}} = \frac{\mathcal{P}}{E(z)}.
    \label{eq:bao2}
\end{equation}
In the case of weak redshift-space distortions an isotropic analysis measures an effective combination of Eqs. (\ref{eq:bao1}) and (\ref{eq:bao2}) and the volume averaged distance $D_{V}(z)/r_{s}$, with
\begin{equation}
    D_{V}(z) = [z(1+z)^{2}D^{2}_{M}(z)D_{H}(z)]^{1/3}.
\end{equation}
Assuming all five parameters free, and without using external priors, we get the results exhibited in Table (\ref{table:tabla2}) for Model I.
\begin{table}[htbp!]
\begin{ruledtabular}
\begin{tabular}{lcccc}
\textrm{Parameters}&
\textrm{Pantheon}&
\textrm{BAO}&
\textrm{CMB}&
\textrm{Pantheon+BAO+CMB}\\
\colrule
 $h$ & $0.6536 \substack{+0.1694\\-0.1503}$ & $0.6975 \substack{+0.1307\\-0.1434}$ & $0.6553 \substack{+0.0006\\-0.0007}$ & $0.6490 \substack{+0.0005 \\-0.0005}$ \\ 
 $c^{2}$ & $0.5403 \substack{+0.2992\\-0.3372}$ & $0.4732 \substack{+0.3243\\-0.2994}$ & $0.5514 \substack{+0.0092\\-0.0092}$ & $0.5950 \substack{+0.0280\\-0.0130}$ \\ 
 $\epsilon_{0}$ & $0.4893 \substack{+0.3213\\-0.3015}$ & $0.5306 \substack{+0.3005\\-0.3233}$ & $0.5330 \substack{+0.0120\\-0.0120}$ & $0.5869 \substack{+0.0093\\-0.0110}$\\
 $z_{s}$ & $-0.5029 \substack{+0.0385\\-0.0431}$ & $-0.4504 \substack{+0.3037\\-0.3330}$ & $-0.6607 \substack{+0.0055\\-0.0048}$ &  $-0.6008 \substack{+0.0063\\-0.0029}$\\
 $\Omega_{\mathrm{k}}(0)$ & $-0.0055 \substack{+0.0015\\-0.0018}$ & $-0.0066 \substack{+0.0057\\-0.0049}$ & $-0.0014 \substack{+8.7\times 10^{-6}\\-6.5\times 10^{-6}}$ & $-0.0014 \substack{+8.4\times 10^{-6}\\-8.4\times 10^{-6}}$
 \end{tabular}
\end{ruledtabular}
\caption{\label{table:tabla2} Best fit values for the cosmological parameters of Model I using the Pantheon, BAO, CMB samples and their combination.}
\end{table}
The summary of the posterior probabilities are shown in Fig. (\ref{fig:constantc}).
Using the same samples of data we obtain the best fit values of the cosmological parameters for Model II as shown in Table (\ref{table:tabla3}).
\begin{table}[htbp!]
\begin{ruledtabular}
\begin{tabular}{lcccc}
\textrm{Parameters}&
\textrm{Pantheon}&
\textrm{BAO}&
\textrm{CMB}&
\textrm{Pantheon+BAO+CMB}\\
\colrule
 $h$ & $0.5832 \substack{+0.1955\\-0.1290}$ & $0.7798 \substack{+0.0828\\-0.1888}$ & $0.6492 \substack{+0.0008\\-0.0009}$ & $0.5440 \substack{+0.0110 \\-0.0059}$ \\ 
 $c^{2}$ & $0.5065 \substack{+0.0071\\-0.0044}$ & $0.7384 \substack{+0.1565\\-0.1486}$ & $0.7552 \substack{+0.0071\\-0.0071}$ & $0.9430 \substack{+0.0210\\-0.0180}$ \\ 
 $\epsilon_{0}$ & $0.2044 \substack{+0.1462\\-0.1168}$ & $0.4904 \substack{+0.3358\\-0.3276}$ & $0.4947 \substack{+0.0039\\-0.0031}$ & $0.4313 \substack{+0.0074\\-0.0110}$\\
 $\beta_{1}$ & $0.4938 \substack{+0.0042\\-0.0066}$ & $0.2516 \substack{+0.1620\\-0.1606}$ & $0.2465 \substack{+0.0052\\-0.0052}$ & $0.3310 \substack{+0.0110\\-0.0048}$\\
 $\Omega_{\mathrm{k}}(0)$ & $-0.2404 \substack{+0.0772\\-0.0298}$ & $-0.0705 \substack{+0.0275\\-0.0461}$ & $-0.0150 \substack{+0.4\times 10^{-5}\\-0.5\times 10^{-5}}$ & $-0.1498 \substack{+0.0002\\-0.0001}$
 \end{tabular}
\end{ruledtabular}
\caption{\label{table:tabla3} Adjusted values for the cosmological parameters of Model II.}
\end{table}
The summary of the posterior probabilities are given in Fig. (\ref{fig:variablec}).
\end{document}